# Primer on Detectors and Electronics for Particle Physics Experiments

Alexander A. Grillo
Santa Cruz Institute for Particle Physics
University of California Santa Cruz

*Abstract*
This primer is a brief introduction to the technologies used in particle detectors designed for high-energy particle physics experiments. The intended readers are students, especially undergraduates, starting to work in our laboratory. The references will provide much more information about these topics and may be as useful, if not more, than the primer itself.

## I Background of Particle Physics Scattering Experiments

Physicists have made the most progress over the past 100-plus years at understanding the nature of the sub-atomic world and to some extent how it is connected to the cosmos by conducting experiments that scatter one type of particle off of some type of matter, be it molecule, atom or other particle. The experiment suggested by Ernest Rutherford in 1909 aimed alpha particles decaying from unstable radon towards a gold foil. His finding that some projectiles unexpectedly scattered into large angles gave the first indication that the atom was not a ball of nearly uniform density but rather included a hard core, what we now call the nucleus. To probe deeper inside the nucleus and then deeper inside the constituents of the nucleus, neutrons and protons, better beams of projectiles were needed with better collimation, more intensity and most of all higher energy. Given the dual particle-wave nature of objects at the quantum scale, higher energy projectiles yield shorter wavelengths, which allow probing smaller dimensions. This is analogous to moving from visible light to ultra-violet to x-ray wavelength microscopes to "see" smaller and smaller objects. The need for these higher energy beams was satisfied with the invention of particle accelerators, cyclotrons, synchrotrons, and linear accelerators. The projectiles of choice were electrons and protons, primarily due to the ease of obtaining such bare particles. These scattering experiments through the 1970s were of a fixed target type, just like Rutherford's. The projectile beam was aimed at a target, which was at a fixed position in the laboratory. The objective of the experiment was to detect the scattered projectile and possibly any new particles that were created in the scattering event. Due to the famous Einstein relationship $E = mc^2$, inelastic scattering events have the potential to convert some of the energy of the incoming projectile into new matter, i.e. particles. In this way many new particles were discovered, possibly adding confusion at first but leading the way to a deeper understanding of this part of nature. Later, physicists learned how to make more exotic beams of some of these particles which had been created in these scattering reactions, like beams of pions, photons and even the nearly impossible to detect neutrino.

The energy available to make new particles or probe smaller distances in these experiments is fixed by the energy of the projectile-target system. If both the target and projectile particles can be accelerated to high energy, the energy available in the center of mass of the system is greatly enhanced. In the 1960s, colliding beam machines started to make an appearance in which two beams of particles are aimed at each other. These machines present many new problems to the accelerator designers but there are also new difficulties for the detector designers. This primer will introduce the most common types of detectors used in scattering experiments and the electronics required to make them work. The different design issues for detectors for fixed target experiments and for colliding beam experiments will be discussed, with an emphasis on the latter since most present-day scattering experiments are that type.

## II Types of Particle Detectors

As stated above, all particle scattering experiments, both fixed target and collider types, require particle detectors to find and possibly identify all the particles emerging from a scattering event. A simple fixed target experiment might scatter a projectile electron off a target of protons (often a container of liquid hydrogen) and simply measure the momentum and angle of the scattered electron. This was the method used to originally measure the size of the proton. More complex experiments create inelastic collisions in which many new particles are produced, all of which must be located and identified in order to analyze the event and understand the dynamics of the interactions.

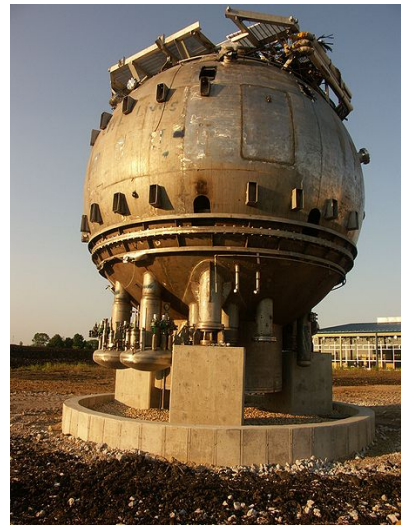

The first particle detectors were visual in nature. The Wilson Cloud Chamber relied on charged particles leaving a trail of small droplets in a super-cooled gas. The trail, called a track, was visible and could be photographed to make a record of the event. Bubble chambers and streamer chambers were significant advances of this principle in that photographing an event and clearing the tracks for the next event could be synchronized to an accelerator beam. All of these visually oriented detectors relied upon photographing the tracks and then measuring the track trajectories in the photographs. The chambers were usually enclosed in a magnetic field such that the charged particles would bend. The measured radii of curvature determined their momenta and the direction of the bend determined the sign of their electric charges. Such a bubble chamber photograph is shown in Figure 2.

**Figure 1: Fermilab's 15-foot Bubble Chamber**

All modern particle detectors rely on the interaction of the particles with material to create an electronic signal, which can then be fed to a computer for analysis. A whole series of gas detectors that accomplish this have been invented and employed over many years. The gas is contained in a chamber in which wires are strung. Positive and negative voltages are applied on the wires in a particular arrangement to create electric fields inside the chamber. When charged particles traverse the chamber and ionize some of the gas (i.e. knock out atomic electrons creating positive ions), free electrons are collected on the anode (with positive voltage applied) wires and positive ions on the cathode (with negative voltage applied) wires. The collected charge can then be amplified and transmitted to a computer thus avoiding the photographing and measuring steps. Some examples of these gas detectors are called spark chambers, multi-wire proportional chambers, drift chambers and even time projection chambers. Figure 3 shows the electrical components of a planar wire chamber. There are two outer wire planes forming the cathode layers and the inner plane of anode wires to collect the negative charge which is then amplified by the electronics. All three wire planes would need to be enclosed in a single outer frame to keep the wires taut and both faces covered with some thin material like mylar to hold the gas volume. This would form one measurement plane, defined by the plane of anode wires, measuring the position of the track in the coordinate perpendicular to the direction the anode wires are strung.

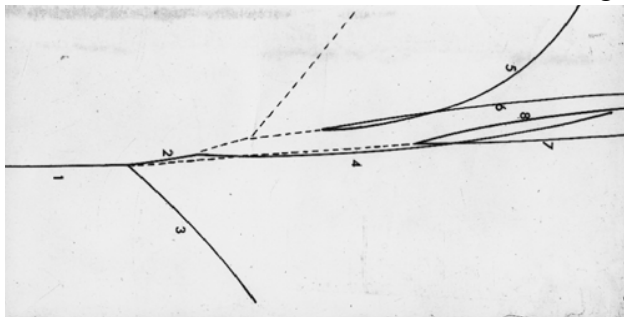

**Figure 2: Bubble Chamber Photo. Charged tracks curve in the magnetic field. Dotted lines are computer reconstructions of neutral tracks.**
**(CERN Bubble Chamber Web Site)**

Another geometric form of wire chambers has become popular in the last few decades. These are built by inserting an anode wire through the center of a long rigid tube of several millimeter diameter. The inside of the tube is metalized to form a cathode surface and the anode wire is stretched very taut. The tube is filled with an appropriate gas and forms one cell of what is called a straw-tube detector. Many, many of these straw tubes are stacked to form a three-dimensional array with the tube lengths running roughly perpendicular to the direction of the impinging particles; the array is thus capable of measuring the full three-dimensional trajectories of charged particles. These have become popular for collider detectors partially because the tubes can be stacked to fit the curved contour as a barrel around a colliding beam interaction region.

A Geiger counter, often found in an undergraduate lab, is an extremely simple type of gas-wire chamber with a metal tube forming the cathode and a single anode wire held in the middle of the tube. Every time a particle passes through the gas volume in the tube, ionized electrons are captured by the anode wire causing the Geiger counter to sound a click. This type of counter is only useful to count radiation activity, for example to detect the presence of radioactive material.

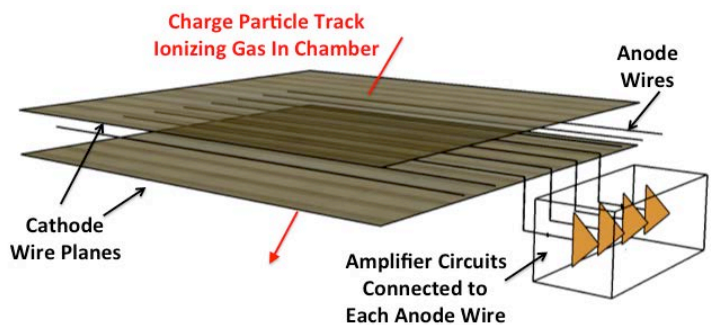

**Figure 3: Wire Chamber Schematic.**
**Anode Wires and Cathode Wire Planes form electric fields in chamber. Particle traverses along Track and Ionizes Gas; Ionized Electrons collect on Anode Wire and are amplified by Amplifier Circuit.**

Another type of detector technology employs special materials called scintillators, which emit light when traversed by a charged particle. This light can be collected and turned into an electrical signal by a device called a photomultiplier. These photomultipliers originally employed a tube technology but recently solid-state devices have become available. With the exception of the time projection chamber, these wire chambers and scintillation counters are usually arranged as multiple planes such that each plane measures the particle position in the one dimension in the plane of the detector perpendicular to the direction of the wires or the scintillation strips. Pairing two planes at a rotated angle yields three coordinates of the track (including the position of the detector planes along the track). Thus, utilizing several pairs of these detector planes inside of a magnetic field allows the measurement of the track trajectory and the calculation of the track momentum and charge sign directly by a computer.

A new type of detector has been developed in the last few decades, which makes use of the advances in semi-conductor integrated circuit technology. Silicon is the standard material but other more exotic materials have been tried. These detectors are fabricated by implanting diodes in a silicon wafer and reverse biasing the diodes. When a charged particle (electron, proton, muon, etc.) traverses one of these diodes, electron-hole pairs are produced with the electrons swept to the anode and the holes to the cathode of the diode. This is similar to the case of free electrons and ions formed in a gas chamber, however, in a silicon lattice the relative energy levels of the valence and conduction bands shifted by the electric field in the diode depletion region allow the charges to move. This small amount of charge, a few femto-Coulombs (fCs), can be amplified and registered as the presence of a particle at the location of that diode. The wafers can be patterned using standard photolithography techniques into diode strips, called micro-strip detectors, mimicking the wires of gas chambers described above to measure one coordinate of the particle's position, or patterned into small rectangular cells called pixel detectors to measure two coordinates of the particle's position. Figure 4 shows a cut-away cross-section view of one of these silicon-based detectors. One of the big advantages of these silicon detectors is that the fabrication process allows the size and spacing of these diodes to be of order tens of microns, whereas wire spacing in chambers cannot be less than a few millimeters and scintillator segmentation not better than a few centimeters. (Pitch is a common term defined to be the distance from the center of one detector element, silicon strip, wire or scintillator segment, to the next.) Thus, the position accuracy of the silicon detectors can exceed that of wire chambers by two or three orders of magnitude and that of scintillators by at least four. Several planes of these silicon detectors can be arrayed just like planes of gas chambers or scintillation counters to measure the trajectories of charged particles.

The various types of detectors discussed above are only sensitive to charged particles but allow the momentum and the sign of their electric charge to be measured by measuring their curvature in a magnetic field. To measure the energy rather than the momentum of a particle, detectors called calorimeters are used. These detectors sandwich several layers of some type of charged particle detector, like the ones described above, between alternating layers of some dense material like lead or iron. The high-energy particles interact with the dense material creating more particles, many of them charged, which share the energy of the original particle. Providing enough layers to stop the initial particle and all of the secondary particles allows the energy of the initial particle to be measured by measuring the amount of total charge deposited in the layers of active detector. These calorimeters are also sensitive to neutral particles in that the neutrals also interact with the dense layers of material creating charged particles detectable in the alternating layers of charged particle detectors. Some modern calorimeters replace the gas in the chambers with a dense liquid such as liquid argon to enhance the absorption of the energy of the secondary particles.

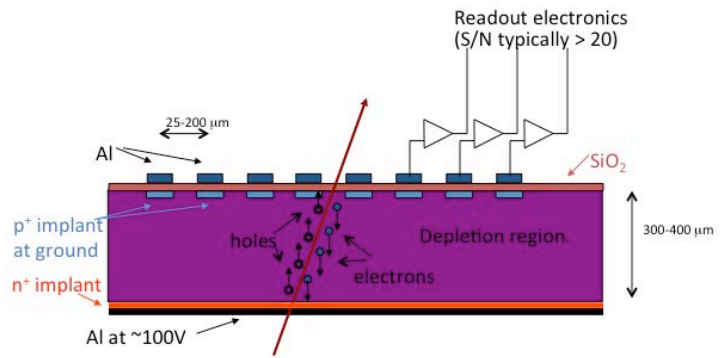

**Figure 4: Silicon Sensor Schematic Cross-Section.** The amplifiers are in external integrated circuits typically wire bonded to the sensor wafer. S/N is signal-to-noise ratio. Al is aluminum conductor and $SiO_2$ an oxide insulator.

In the following sections, a distinction will be made between "sensor" and "detector". A sensor will refer only to the material through which the particles traverse creating some sort of detectable signal, e.g. the silicon wafer patterned with diodes or the chamber filled with gas and strung with wires. A detector will refer to a more complete detection system, at a minimum including the sensor and its readout electronics. There will, however, still be multiple uses of this term "detector" in that as experiments have grown, they have needed to combine several different detectors or detector types for one experiment, and so the term detector can refer to the combination of all detectors for one experiment such as the ATLAS detector or for the individual parts of that combination such as the ATLAS Pixel Detector or the ATLAS Liquid Argon Detector. And each of those detectors can and will be made up of many separate sensors plus electronics systems, each of which could be called a detector. The context of each usage will hopefully make the meanings clear.

A more comprehensive discussion of particle detector types is covered in *Particle Detectors* by C. Grupen [1]. Also, the book by H. Spieler, *Semiconductor Detector Systems,* is an excellent resource for detectors and their electronics using semiconductor sensors [2]. The electronics associated with all these different detectors must sense very small signals, amplify them to a usable level with a sufficiently good signal-to-noise ratio (the ratio of the amplified signal size to the combined amplified input noise plus the noise introduced by the amplifier circuit), and then condense the data into some form usable by analysis computers.

While the materials used to sense particles and the corresponding electronic signals are nominally the same for fixed target experiments and collider experiments, the necessary arrangement of the sensors and resulting mechanics create some very different design criteria for the sensors and their readout electronics. Conservation of energy and momentum forces all the products of a fixed target scattering experiment to fly into a forward cone in the lab, while products from a collider scattering experiment, that is where the center of mass frame is the same as the laboratory frame, emerge at all angles in a $4\pi$-steradian solid angle in the lab. The density of particles per unit solid angle is then greater for fixed target experiments, possibly putting more difficult requirements on position resolution of the sensors, but the limited area of the full detection system normally makes mechanical supports, access to electronics for repair and routing of electrical and cooling services all much easier since the outer perimeter of the detection system is open and accessible.

Some experiments are designed to study only one specific reaction or test one specific concept. These experiments, which will require a special purpose detector, may not need to capture all the particles produced in each event possibly simplifying their design. This can be true for both collider or fixed target experiments. If the purpose of an experiment is to examine all the products of the scattering, then its detector must cover as much of the full $4\pi$ steradians in the interaction's center of mass frame as possible. This is true for both collider and fixed target experiments. The difference comes from the translation from the center of mass frame to the lab frame. The $4\pi$ detector at a collider is a $4\pi$ detector in the lab, i.e. at the collision point of the two beams. The $4\pi$ detector for a fixed target experiment need only cover a region forward of the target in the lab. As the cost of building colliding beam facilities has increased and the reactions of interest have become more rare, forcing experiments to collect data for many years, the desire to make the most of each interaction has made full $4\pi$ geometry detectors more attractive. Certainly, each colliding beam facility must have at least one if not two such general purpose (i.e. examine everything) detectors along with possibly one or two special purpose ones. The "4π geometry" in the lab of those collider detectors complicates the design issues, as we will see in the next sections.

## III Integrated Particle Detectors for Experiments

As just stated, most but not all detectors for collider experiments are designed to capture as much of the complete solid angle surrounding the interaction point as possible. This is important in order to detect all the particles produced in the scattering event and thereby reconstruct what occurred. Furthermore, some particles likely to be produced in these interactions, namely neutrinos, do not readily interact with matter and require tons of material to detect, much more material than is feasible for a general-purpose collider detector. Therefore, these particles are not recorded by the collider detector system. Instead, missing energy in the event is attributed to neutrinos, making the full coverage, typically referred to as hermeticity, essential so as not to lose any detectable particles and incorrectly attribute that loss to an undetectable neutrino. The ATLAS detector built for experiments at the Large Hadron Collider (LHC) at the CERN laboratory in Geneva, Switzerland is a good example of such a hermetic full-coverage detector [3].

A cut-away drawing of this detector is shown in Figure 5. The two colliding beams enter through the tube on the far left and a similar tube on the right that cannot be seen and collide at the very center of the detector. The full detector is made up of many sub-detectors, each of which surrounds the interaction point and serves to detect a certain type of particle. The Inner Detector, closest to the interaction point, which includes the Pixel Detector, the Semiconductor Tracker (SCT) and the Transition Radiation Tracker (TRT), detects charged particles and resides inside a 2 Tesla solenoidal magnetic field. Each of these three detectors consists of several concentric barrel and end-cap layers to allow determination of the particles' momenta and charge sign by measurement of the radii and direction of curvature of their trajectories. The Pixel and SCT Detectors are examples of silicon detectors described in the previous section, the latter using silicon micro-strip sensors. The TRT employs gas straw-tube technology. Next the Liquid Argon Calorimeters (LAr) measure the energy of electrons, positrons and gammas (photons). Lead sheets are used to force these low-mass particles to interact and electrodes in the liquid argon collect all the charge deposited in that medium. Hadrons (protons, neutrons, pions, etc.) deposit very little energy in the LAr Calorimeters but are stopped in the multiple layers of iron and plastic scintillator sheets of the Tile Calorimeters which measure the hadron energy in the same way. Finally, the Muon Spectrometer detects muons, which are the only particles that emerge from the calorimeters as all others have been stopped. The Muon system is called a spectrometer because it includes large toroid magnets which again bend the charged muons allowing their momenta and charge sign to be measured by their curved trajectory with precision wire chambers. There are some other technologies used in the muon system to facilitate a function called triggering but the entire topic of triggering and the required trigger electronics are beyond the scope of this primer. The enormous size of the detector (See the small 6-foot tall persons in Figure 5 for scale.) is required in order to capture all the energy from the reactions and to provide sufficient travel distance through the magnetic fields to measure the momenta of the very high-energy particles.

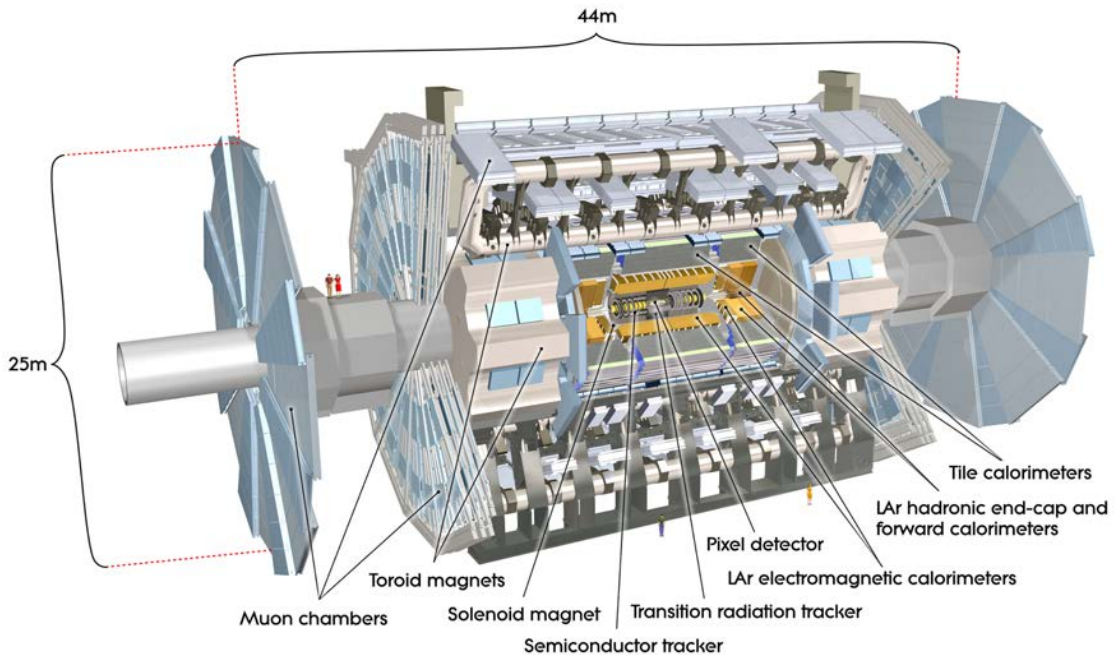

**Figure 5: The ATLAS detector. (**ATLAS Experiment © 2014 CERN**)**

Figure 6 is a drawing of the detector for the Heavy Photon Search (HPS) experiment at the Jefferson Lab in Virginia, USA [4,5]. This is both a fixed target experiment and a special purpose detector but it still contains most of the necessary components of a general purpose detector. A beam of electrons enters the detector area from the left and strikes a thin gold target. Without explaining the physics being studied, the primary product of the interactions is an electron and positron pair in addition to the original beam electron. The detector then must detect these three particles, measuring well their trajectories and energies, but must also detect any other particles produced since that would be a background for the experiment. Even that requirement to detect all particles produced can still be met with detector elements only in the forward direction in the laboratory. This detector also has nearly all the types of detectors as the ATLAS detector. There is a silicon tracker followed by an electromagnetic calorimeter and a Muon detector. The only type missing is a hadron calorimeter. This is because few hadrons will be produced with an electron beam on a gold target but any events with hadrons would simply be discarded without need to measure the energy or momentum of these extra particles. To be honest, however, all these types of detectors do not cover the entire $4\pi$ center of mass region. Since any extra particle in an event forces that event to be discarded, requiring the energy of the three detected particles to equal that of the original beam particle can be used to reject events with any extra particles.

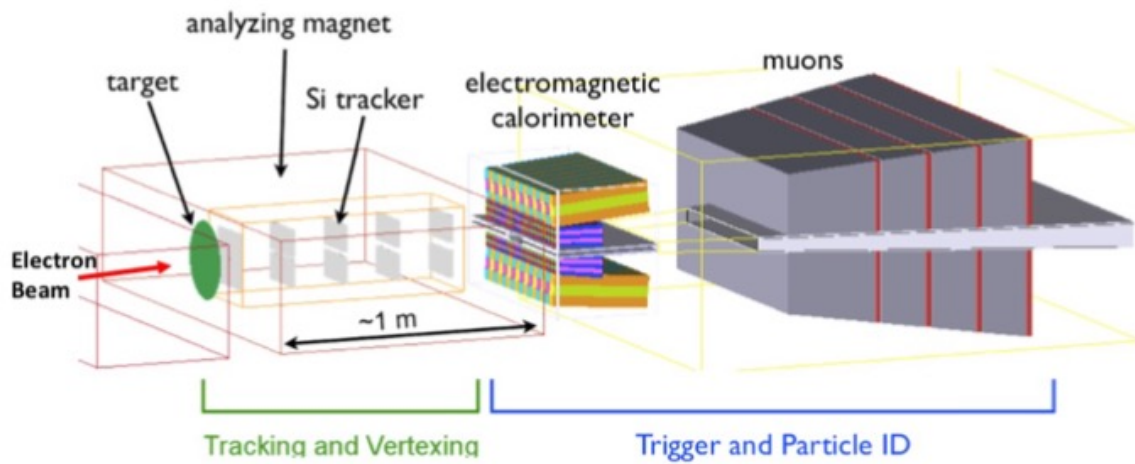

**Figure 6: The HPS Detector**
**A diagram of the detector components with the beam arriving from the left.**

To visualize how an integrated set of detectors works, Figure 7 shows a computer reconstruction of an event recorded in the ATLAS detector. This is a two-dimensional projection of the tracks in the event looking down the direction of the two colliding beams. In the central tracking detector, many tracks are visible. There are sufficient numbers of silicon detectors and TRT straw tubes in this central tracker to record enough hits for the computer to reconstruct tracks and produce visible representations like the bubble chamber picture of Figure 2, except that in this case, the visual representation is only for the benefit of humans. The computer takes those same track parameters to directly reconstruct the kinematics of each event. Most charged and all neutral particles are absorbed by the calorimeters with only the four muons escaping to be recorded by the muon spectrometer at the outer most radius. Keep in mind that there is also a third dimension to these events. The tracks have some extent into that dimension as well.

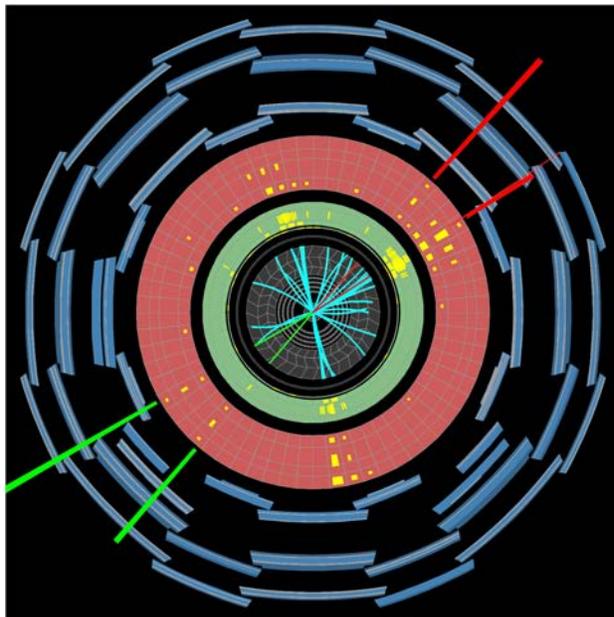

**Figure 7: Computer Reconstructed ATLAS Event.
Charged tracks in the Inner Detector;
Energy deposited in the two Calorimeters;
Four tracks in muon chambers.**
(ATLAS Experiment © 2014 CERN)

Note that the CMS detector also at the LHC is a second example of a nearly $4\pi$ hermetic detector while the ALICE and LHCb detectors at the LHC are designed for more special purpose experiments [6,7,8]. Some of the requirements discussed here may not apply to those latter two detectors.

## IV Design Issues for Detectors

The design of detectors for high energy particle physics experiments including which technologies to use are all in the domain of the creativity of the experimenters. There are some issues which are common to any of these experiments but some are specific to or more extreme for collider experiments than for fixed target experiments. The picture of the ATLAS detector demonstrates one of the design issues for detectors for collider physics experiments, reliability. With the possible exception of the outer most sub-detector (in this case the Muon Spectrometer), detector components will be buried deep inside the massive system. Accessing those components for repair or replacement involves opening up the entire apparatus, a task requiring many months of downtime and involving risks to other fragile components. The experiments are typically designed to run for up to 10 years and, therefore, all the components must be designed to operate for at least that long without significant failure.

The other issues for these detectors include minimizing material, low power because of limited space for services, extremely limited room for both power cabling and cooling plumbing, potentially high data rates and, perhaps the most onerous, radiation tolerance since the very nature of the application requires operation in a high radiation environment. These issues can be compared to detectors for typical fixed target experiments.

Again, consider the Heavy Photon Search experiment as the counter example. Figure 8 is a picture of the HPS Silicon Tracker, three planes of silicon strip detectors. The services attach at the side and top of the planes with no issue of services covering the particle trajectories. Care to minimize material for services is not an issue and servicing parts of the detector is a matter of days, not months.

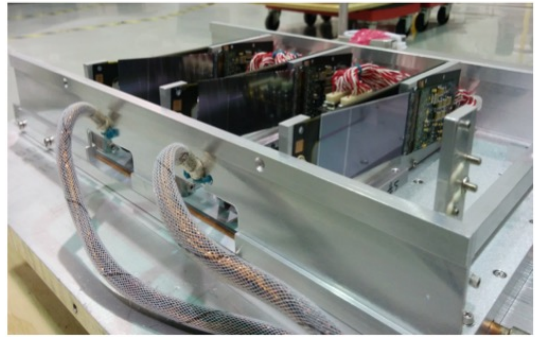

Figure 8: The HPS Si tracker Beam impinges from the right. Note services attach from left and top of silicon strip detectors without blocking the active tracking volume.

Each of the issues for detector systems will be considered in the following sub-sections. Electronics used inside these detectors have special requirements and they will be discussed separately in section VI.

*IV-a Radiation Tolerance*

The particles created in these experiments, which the detectors are designed to detect and measure, are forms of radiation that can damage many types of material. Certainly, the sensor areas of the detectors must be able to withstand this radiation during their lifetime, but the $4\pi$ nature of most collider detectors results in the support components, e.g. mechanical supports, cables, cooling pipes and electronics, having to also survive the expected radiation. To some extent this is true for all experiments due to stray radiation produced as a byproduct the interactions. Experimenters have been surprised by several types of material degrading after exposure to large doses of radiation. Teflon used as insulation for cables has turned to powder. Wires inside of chambers with some exotic gas have developed corrosion over time. Experimenters have learned that any material planned for the inside of one of these detectors that will be exposed to radiation, either particles that are to be studied or peripheral radiation created during the interactions, must be tested for radiation tolerance. Beyond these general statements, it is impossible to cover all the varied detector technologies here. The one exception is the silicon sensor that shares many of the same radiation sensitive issues as the readout electronics. For that reason, more about radiation tolerance of silicon sensors will be included in the electronics radiation section VI-a. The amount of radiation exposure depends upon many factors such as the type of beam, beam intensity, and distance from the interactions. A few representative examples are given in Table 1 of section VI-a, which includes more information about radiation issues.

*IV-b Minimal Material*

When particles traverse through material, they lose energy and often scatter. These interactions change the tracks' curvatures which confound the measurements of the particles' trajectories and hence their momenta and energies. The sensor materials, e.g. silicon wafers or chamber gas, are naturally low mass but are still unavoidable material in order for detection to take place. The mechanical supports, cabling, cooling pipes and electronics, however, represent extra material, which further degrades the performance of the detectors if particles must pass through them. As the volume to be covered by detectors increases, it becomes unavoidable that some support structures or other services shadow some of the sensor volume with the extreme case of a $4\pi$ collider detector where sensors and support share the exact same volume. The ideal detector then would consume zero power, be light enough to be supported by gossamer strings and transmit its data wirelessly to the data acquisition system. No one has yet invented such an instrument but experimentalists are always trying to minimize the material inside these complex detectors.

*IV-c Low Power*

This issue is mostly a consequence of the material issue discussed above. Power dissipation in a detector requires cables to deliver the power and a method to remove the power, usually cooling pipes. Both represent material inside the detector volume. A second reason for this issue is limited space for these services and if the detector is supposed to be hermetic, all of these services require access points, which likely will break the hermeticity.

Most of the power consumed by a detector is by the electronics mounted on or very near the sensors. Scintillation sensors consume no power; wire chambers themselves consume very small amounts of power needed to provide required voltages on the anode and cathode wires; silicon sensors also consume a very small amount of power in the form of leakage current in the reverse biased diodes, however, this leakage current can grow to be significant after the sensors have been exposed to a substantial amount of radiation. More about this in section VI-a.

*IV-d High Data Rates*

The successful advancement of particle physics in the last 60 years has pushed the experiments to look for more rare processes as well as to search at higher and higher energies. As a consequence of this, the detector systems have grown in size, increasing also the number of readout channels. (The ATLAS Pixel detector has 80 million channels and the SCT detector has 6.3 million channels.) The beam intensities have increased enormously as well. The LHC collides two proton bunches every 20 ns. This has resulted in large increases in the amounts of data and data rates to be transmitted out of the detectors. This will be discussed more in the electronics section VI-d.

*IV-e High Reliability*

Lastly, we return to an issue that was raised at the very beginning, namely reliability. As mentioned already, colliders are typically designed to operate for at least 10 years with only minor repairs. This is due in part to the large expense in building one of the detectors and the beam facility, but there is much to be learned from the experiments' use of them and it usually takes many years to fully exploit the potential of the investment. In fact, it is often the case that after 10 years more can be learned with further operation of the detector and collider but upgrades are required to enhance performance, capitalize on advances in technology and replace some worn out parts. Radiation damage often limits the lifetime of some components. Therefore, 10 years is typically a convenient lifetime specification. Certainly, it has been the case that after 10 years new technology allows significant improvement in performance of some components.

During this ~10 year lifetime, any component failures can cause serious problems. Opening one of these detectors can take several months, possibly as long as a year. This down time is very expensive in that beam time is lost, operations personnel are still on payroll and experimenters are delayed in their work. For those detectors that subtend a large solid angle with each detector type wrapped around those closer to the interaction point, servicing an inner detector requires opening and often removing all the outer ones. The work is tedious, in a confined space underground, sometimes requiring outer sections to be lifted out to the surface. Many components like cable and plumbing connections are fragile having been designed to minimize material inside the detector such that damage can be done in the process of fixing something else. And lastly, after several years of operation, the inner components are probably activated, i.e. they emit radiation, due to a change in the nuclear structure of their material from the constant exposure to radiation. This requires special handling of all activated components because the activated material can be dangerous for humans to handle.

To ensure high reliability, careful engineering practices are exercised with quality assurance procedures in place and fully exercised. Lifetime tests are performed on as many components as possible, especially for components which do not have published industry lifetime test results. Since there are very few if any moving mechanical parts inside such a detector (pumps and fans, etc. are always located outside of the actual detector area where they can be serviced more easily), failures due to radiation damage and corrosion are the primary "wear out mechanisms". There are special issues regarding reliability and lifetime of electronic components. These will be discussed in sub-section VI-e.

## V Electronics for Particle Detectors

The electronics associated with particle detectors can be separated into two groups: on-detector electronics that are mounted in close proximity to the sensors and off-detector electronics located many meters, perhaps a few hundred meters, away from the sensors. The main goal of the on-detector electronics is to amplify the very small sensor signals, integrate and shape the signals over a relevant time period, and provide robust signals to be transmitted off detector. The off-detector electronics can then perform much more complicated data processing such as track reconstruction, cluster analysis of adjacent calorimeter channels and decisions about the relative interest of a particular event, i.e. is a particular scattering event worth recording on the storage media for later physics analysis. Linking the on-detector and off-detector electronics will be some form of data transmission, now typically optical. Given that the off-detector electronics are located outside the immediate detector area, usually in a side service cavern or surface building, they do not reside in what would be considered an especially harsh or extreme environment. Therefore, we will focus only on the on-detector electronics.

In addition to the varied types of sensors developed for particle detection, there are a few different methods for electronically processing the data depending upon what information is desired. The simplest information is "did a particle traverse this sensor?" This so-called hit or no-hit information can be very useful if each sensor is very finely segmented. For example, the silicon micro-strip sensor mentioned above can form silicon diode strips with a pitch as small as 50 $\mu$m. Recording which strip has a hit then can identify the position of a track to very high accuracy. Potentially more information can be obtained from the sensor. The actual amount of charge deposited in the sensor can also be recorded. For a calorimeter, this is essential for its operation.

The amount of charge deposited in a calorimeter is a measure of the energy of the detected particle. Electronics for these detectors must record a full analog value for each sensor channel.

Figure 9 shows a block diagram for a typical front-end circuit providing the simple hit/no-hit information. One and possibly two stages of amplification are required followed by an integrator/shaper block and then a comparator to separate a real hit from noise. Finally, there is some translation to a digital signal. Following these front-end blocks, there will likely be buffering of data, building some kind of output packet and transmission protocol. For applications where analog information is required, the comparator and digital translation stages will be replaced with either analog buffering and some sort of analog transmission or more likely these days with analog-to-digital converters (ADCs) built into the on-detector ICs (integrated circuits) and digital buffering and data transmission. The on-detector electronics also typically include some monitoring functions like tracking the temperature and humidity of the detector.

It is not possible to discuss in detail in this short work all the varied types of data processing now being employed in particle detectors. The availability of very dense integrated circuit technologies has made possible very complex on-detector circuitry, and more functionality has been moved from off-detector to on-detector. ADCs are a good example of this. In the past, all ADCs had to be located off detector because of power and space requirements. Moving more electronics on detector removed the necessity to transmit analog signals over long distances to the off-detector racks. This has reduced noise and signal attenuation allowing detection of much smaller sensor signals. It also afforded the possibility of greater compaction of data before transmission, possibly reducing the transmission bandwidth requirements. This has been essential as the channel counts have grown. There is, however, always a compromise to be sought between the complexity of the on-detector electronics and the size of the data transmission system, especially as regards reliability and material inside the detector volume. The issues of signal vs. noise and pulse shaping will be expanded in section VI-b.

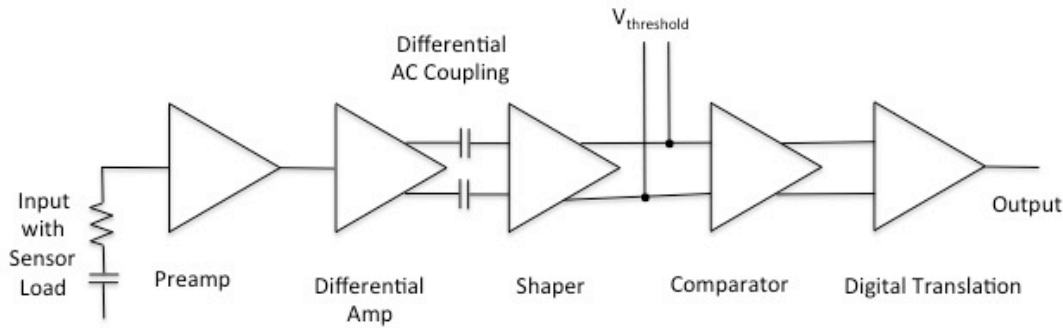

**Figure 9: Block Diagram for a Typical Front-end Circuit for Hit/No-Hit Readout**
The preamplifier, amplifier and shaper circuits amplify the signal and possibly integrate over the expected charge collection time of the sensor. AC coupling blocks any offset or leakage DC currents. The comparator discriminates signals larger than $V_{threshold}$ from lower level noise.

## VI Requirements

Several issues specific to particle detectors especially those designed for collider experiments have been discussed above in section IV. Those will be expanded here in the next several sub-sections as they apply to the detector electronics. Many of the requirements are common to both fixed target detectors and collider detectors. This will be noted in each sub-section.

In general electronics can be of various technologies, vacuum tubes, discrete solid-state components or integrated circuits made of silicon, gallium arsenide and other materials. Indeed, particle physics has grown hand in hand with the electronics technologies. Early systems in the 1940s and 1950s used vacuum tubes. These were replaced by discrete semi-conductors and later by ICs. The advances in the electronics industry have afforded many of the advances in particle physics. A detector like ATLAS could not have been built without the readily available IC foundries, which can fabricate ICs custom designed for the detector applications. Most requirements for these detectors can now be met by commercially available integrated circuit technologies with custom designs.

There are two other points to keep in mind as the various requirements are considered. All of the sensor types used for particle detection involve collection of charges released in the sensor material by the impinging particle radiation. This collection of charges requires voltages much larger than that powering the attached readout electronics. Silicon sensors typically require a few hundred volts to possibly 1 kV. Other sensors such as wire chambers or liquid argon calorimeters require several thousand volts. These high sensor voltages are always blocked from the readout electronics with capacitive couplings but the presence of such voltages often has implications for system design, for example appropriate voltage protection devices. The other point is that many if not most of these detectors are positioned inside of a strong magnetic field, which normally prevents the use of any device made of magnetic material, like magnetic core inductors for filters or for DC/DC buck converters. Very recently, since the original version of this paper, exotic materials whose magnetization saturates only at very high magnetic fields have become available as replacements for common ferrites in commercial inductors. This may facilitate the use of these filters and DC/DC converters instead of relying only on air-core inductors.

*VI-a Radiation Tolerance*

The particles, which these detectors are designed to study, are exactly the forms of radiation that can be damaging to microelectronics. While there are possibilities in fixed target experiments or special purpose detectors to keep the electronics out of the radiation area, that is impossible with the "4π geometry" collider detectors. However, sometimes the electronics for fixed target detectors must also be exposed to radiation in the interests of keeping the first stage readout immediately at the sensors to reduce noise and optimize data transmission. Also, the beams themselves are a large source of background radiation. Beam designs employ dipole bending magnets and quadrupole magnets to focus the beams on the targets or on the interaction points with collimators to scrape away any off-momentum particles. These collimators do not completely stop everything hitting them resulting in some lower energy remnants entering the experimental area. Neutrons are an especially difficult problem in this regard. All modern collider detectors and most fixed target ones, as well as the accelerators or storage rings that feed beam to them, are buried underground so that the surrounding earth can provide shielding for human safety. The on-detector electronics must live in the radiation environment. As the field of particle physics has advanced, it has become necessary to study reactions with lower and lower probabilities (referred to as cross-sections). This has forced the beam intensities to increase thereby increasing the radiation exposure of the electronics. There are both long-term permanent damage and instantaneous disruption, often referred to as Single Event Effects (SEE), which must be considered in designing the electronics. An introduction to the effects of radiation on electronics will be given here. A more detailed discussion of the effects on semiconductor devices is covered in the book by H. Spieler [2].

The required level of radiation hardness for different collider experiments depends on many factors. Hadron (typically proton or anti-proton) colliders typically produce more radiation than electron/positron colliders because of the nature of the scattering events. Radiation increases with the luminosity of the colliders, which is increasing with each generation as more rare processes are being studied. Most important is the position of the electronics. Readout electronics for an inner most tracking detector will experience much more radiation than that for a calorimeter or muon detector, which are typically located outside of the tracking detectors. As an example, Table 1 shows the expected exposures for two different detectors, BaBar at an electron/positron collider at the SLAC laboratory and ATLAS at a proton/proton collider at the CERN laboratory.

**Table 1: Examples of Required Radiation Hardness
of Two Typical Collider Detectors (From [9,10,11])
For definitions of the units in this table, see the Appendix.**

|  | Total Ionizing Dose | Non-ionizing Fluence (1 MeV neutron equivalent) |
|---|---|---|
| BaBar Inner Silicon Strip Detector Outer Calorimeter | 20 kGy  100 Gy | Negligible  Negligible |
| ATLAS Inner Pixel Detector Outer Muon Spectrometer | 500 kGy  20 Gy | $10^{15}$ $n_{eq}$/cm$^2$  $10^{12}$ $n_{eq}$/cm$^2$ |

With the exception of some passive components such as resistors and capacitors, on-detector electronics are primarily made of ICs. Some dielectrics used in discrete capacitors can be sensitive to radiation and therefore should be qualified before use, however, we will focus attention here on radiation damage to ICs. While there are several different IC technologies these days, they can generally be grouped into two main classes, MOS (Metal Oxide Semiconductor) field effect transistor devices and bipolar junction transistor devices.

MOS field effect transistors or MOS FETs now come in various flavors such as NMOS, PMOS, CMOS, LDMOS. The basic concept of all of these is illustrated by the cross-section view in Figure 10. A voltage is applied between the source and drain terminals and an intermediate voltage level is applied to the gate terminal to control the current flow between source and drain. The voltage applied to the gate terminal creates a field in the region between the source and drain implants just under the gate oxide layer, which controls current flow between the source and drain. Ionizing radiation in the form of charged particles or high-energy gammas can damage the thin gate oxide and form trapped charge sites. These trapped charges will then alter the electric field under the gate oxide changing the current vs. gate voltage relationship. This results in a change to the threshold voltage of the device, i.e. the gate voltage at which the transistor turns on or off. This will affect the speed at which the transistor will turn on or off as well as the gain of the device, and given enough radiation exposure, the transistor can become locked on or off. Even before this occurs, the change in threshold voltage can cause the transistor to leak current, that is, to not be fully off when the gate voltage would normally have produced the off state. This leakage current increases the power consumption of the IC. Another radiation effect is damage to the thicker "field oxide" which covers the chip and provides some added isolation between transistors. Trapped charge in this field oxide can induce current paths between transistors, also increasing the power consumption of the IC.

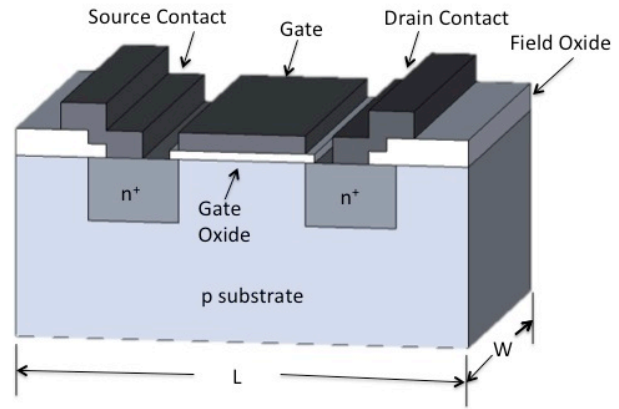

**Figure 10: N-channel MOS FET Cross Section. Conduction channel is between the source and drain implants under the gate oxide.**

Bipolar junction transistors or BJTs, often called just bipolar transistors, operate by a different mechanism and experience different radiation damage effects. A typical bipolar transistor cross-section is shown in Figure 11. The back-to-back n-p and p-n junctions of the device allow current flow through the base terminal to control the current flow between the collector and the emitter. The resulting triode functions in much the same way as an MOS transistor, however, the current flows more directly through the silicon body and is dependent upon the relative dopings of the junctions. Damage to the silicon lattice, called displacement damage, then can have detrimental effects to the transistor performance. Such damage to the silicon bulk is not typically produced by ionizing radiation but rather by non-ionizing radiation (e.g. neutrons, protons, pions). Of course, the charged versions of these particles like proton, $\pi^+$ and $\pi^-$ are also forms of ionizing radiation. Oxides are used to isolate the three terminal areas of the device. Depending upon the details of how the transistor is structured, these oxides can provide unwanted current paths if they are damaged by ionization radiation. Radiation damage to bipolar devices normally results in increased base leakage current, which is lost to the collector-emitter control. That is, the gain of the transistor, which is defined as the collector-emitter current divided by the base current, decreases because the base leakage current is increasing. The effectiveness of the transistor then degrades possibly to the point where it is no longer usable. Thus, bipolar devices are susceptible to damage by ionizing radiation and by non-ionizing radiation. MOS devices are usually immune to non-ionizing radiation damage but will suffer from ionizing radiation.

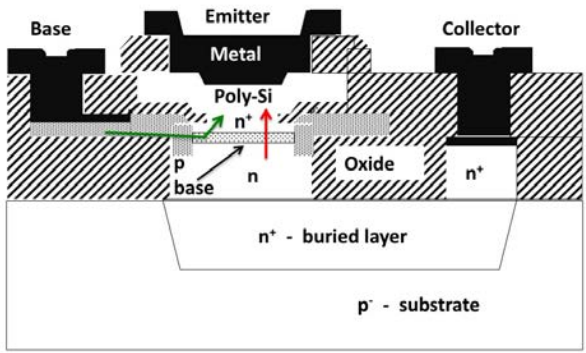

**Figure 11: BJT Cross-section**
**A vertical n-p-n transistor structure.**
**Green arrow tracks base-emitter current, which controls collector-emitter current through thin base layer, tracked by red arrow, and buried layer.**

A brief word about silicon detectors, or more specifically silicon sensors, can be made now. As pointed out earlier, these sensors are fabricated as diodes implanted in silicon wafers such that the diode junctions collect the charge deposited by the traversing charged particles by applying a significantly large reverse bias voltage. In many respects then, they can suffer similar damage to bipolar electronics. Indeed, the most serious damage occurs as displacement damage by non-ionizing radiation. This results in increased leakage current of the reverse biased diodes and larger voltages required to fully deplete the bulk. Effects can also be seen from large doses of ionizing radiation but these effects often appear as charges on the surface of the sensors where oxides are again used as insulators. Careful design of the diode structures and implant doping densities have yielded sensors which can survive large amounts of radiation resulting in sensor lifetimes of up to 10 years in present day colliders [23].

For many years, special highly proprietary techniques were developed by a few IC vendors to mitigate the radiation effects in semiconductors. This was especially true for CMOS technologies. They were aimed at military markets but were also useful for particles physics. This included special "black magic" oxides that somehow did not develop the charge traps which increased leakage currents and changed the threshold voltages. Bipolar devices could be built into more vertical structures, which changed the locations of the oxide isolators reducing their post-irradiation effects. All these techniques provided some level of radiation immunity but not complete immunity, and the proprietary CMOS technologies were very expensive.

As commercial technologies have continued to shrink the device feature sizes, these technologies have by coincidence become more immune to radiation. For example, the thinner gate oxides for MOS devices have resulted in manageable radiation-induced threshold shifts because the electron tunneling rate (a quantum mechanics effect) is sufficient to neutralize the charge traps. Trench isolation between the transistors has reduced the leakage current between transistors. The smaller area of the bipolar base has reduced its cross-section (probability) for hadronic induced lattice defects, however, a potentially bigger help for BJTs has been the introduction of lattice stress in the base region (e.g. silicon germanium technologies), which has greatly increased the current gain such that the radiation induced decrease is not as critical. Still some care must be taken in circuit design for these technologies to be completely acceptable. Leakage currents, especially between structures, remain a concern for technologies relying on field oxide isolation. One method to mitigate this in FET designs is to create an enclosed geometry with one terminal surrounding both the channel and the other terminal thus allowing the electric potential of the surrounding terminal to block leakage currents that could result from field oxide damaged by radiation [12]. Figure 12 shows a representative 3-D sketch of such a geometry.

The values in Table 1, especially for the ATLAS detector, may be slightly misleading because the high total dose and fluence values do not correspond to a relatively high dose rate or fluence rate. These expected levels of radiation are for a 10-year lifetime of the detector. If one takes into consideration the full 10 years with an operating time of roughly 66% of each year, the irradiation rates even for the inner most system are roughly 2 mGy/s (Gy = Gray = 100 rad) and $5 \times 10^6$ $n_{eq}/cm^2/s$ ($n_{eq}$ = 1 MeV neutron equivalent), relatively low rates. The other application typically identified as requiring radiation tolerant electronics is the military. Some of those applications require equally high total dose and fluence tolerance but they typically also have high dose and fluence rates because the expected time of exposure can be as short as a few days to a few seconds. Fortunately, for several reasons, particle physics applications to not have to meet full military requirements.

One consequence, however, of the low dose rates expected, is that low dose rate effects must be examined for the technologies being considered for the particle physics applications. Low dose rate effect (LDRE) is a phenomenon by which the effective damage to the transistor is enhanced if the radiation exposure is at a low rate rather than a high rate. While this may be counter intuitive, it has been observed with some bipolar technologies and most recently by a different cause but with similar complication in a CMOS technology. It, therefore, must be checked. This can complicate the testing of devices because it is impractical to test for a lifetime dose at the expected dose rate if the target lifetime is 10 years. Total dose and total fluence tests are normally run at high rates in order to make the test time short, possibly a few days. Fortunately, these low dose rate effects usually appear at relatively low total dose and fluence, i.e. in a relatively short time of testing at low rates allowing most parts which have no LDRE to be qualified for total dose and fluence at high dose and fluence rates. An example of this effect and such a study of it is described in [19].

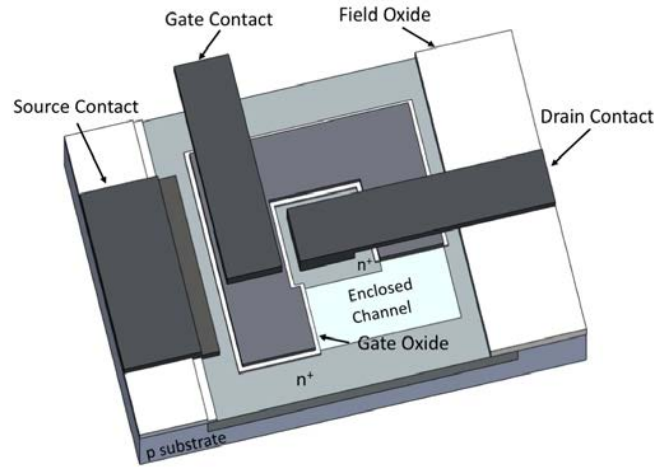

Figure 12: NMOS Transistor with Enclosed Geometry The field oxide has been cut away except under the source & drain contacts and the gate & gate oxide cut away in the front segment to allow a view of the enclosed structure with the drain enclosed by the gate which is in turn enclosed by the source.

One must also be concerned about instantaneous disruption of electronic devices. Even though these disruptions are most common with highly ionizing ion radiation, singly charged minimum ionizing particles can cause disruption. This is especially true as device sizes shrink and the charge required to change state is reduced. There are two types of these disruptions. Single Event Upset (SEU) is when a digital transistor spontaneously changes state, either 0 → 1 or 1 → 0. This is most often caused by an ionizing particle passing through the device, depositing just enough charge to flip the state of one or more transistors. This can cause a change in the state of the logic or the meaning of a memory cell causing the device to malfunction until it is reset. The other type of disruption is called Latch-up. This is caused by the normal existence of parasitic BJTs in the bulk of typical CMOS structures as shown in the schematic cross-section of Figure 13. A heavily ionizing particle passing through the bulk or even a voltage spike on the IC's power input can cause this pair of parasitic transistors to effectively turn on causing a low impedance path through the pair. Such a condition across the supply voltage of an IC can cause permanent damage. In all cases, power cycling the device is required to clear the condition. Latch-up must be avoided at all costs by device design and layout. Single even upsets (SEU) that cause bit flips, which can be restored by re-writing, can be problematic but possibly manageable. Since mitigation techniques for SEUs typically cost power or chip real estate or both, it is usually desirable to evaluate the severity to overall system performance of a particular bit-flip error in order to decide what kind of mitigation technique, if any, to employ.

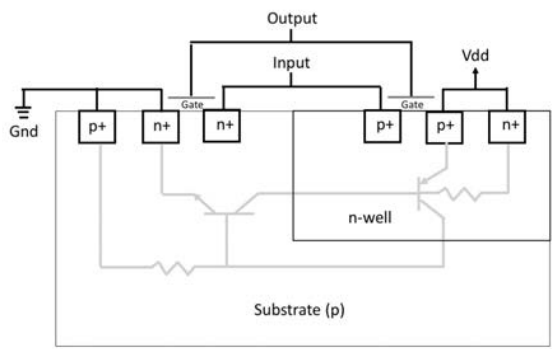

**Figure 13: Schematic Cross-section of Parasitic BJTs in Bulk CMOS**
**The parasitic BJTs shown in gray are internal to the substrate**

To illustrate further, a register that is used to control the operation of the electronics may be critically important such that an internal bit-flip may cause the detector to malfunction and possibly cause some damage. Such a register should be carefully protected, possibly employing triple redundancy voting (a technique that creates three copies of the critical logic or memory cell such that the state of a majority of the three copies will be taken as the state to be used by subsequent logic). Likewise, writing to such a register should be protected with some sort of error correcting code (ECC), a Hamming code is common. A register that is not so critical, perhaps for an on-board digital-to-analog converter to set the threshold for a data comparator, may require lighter protection (for example a more robust latch as a Dual Interlocked Storage Cell [13]) and ECC skipped on writing if read back after write is possible to verify what is written. An example where no correction technique is typically required is in the buffering and transmitting of sensor data. As long as the expected SEU error rate is much lower than the expected inherent error rate of the sensor and readout, there is no need for extra correction. As an example, the silicon micro-strip detector (SCT) for ATLAS has a noise error rate specification of $5 \times 10^{-4}$. Testing of the on-chip pipeline buffer for SEUs found an error rate of $10^{-11}$, clearly negligible compared to the noise rate and so no protection of the pipeline was necessary [14].

In spite of the success in using commercial technologies for detector electronics, extensive testing of each new technology has been absolutely required. This should be expected since commercial vendors do not specify any radiation tolerance and therefore do not guarantee any. This need for testing applies even to two technologies with the same basic commercial specifications but from two different vendors. Often some subtle difference in the three-dimensional layout of the structures in the technology, especially the structure of the oxide isolation layers, can have very significant effects on the radiation tolerance. One concern in this regard is the possibility for a foundry to make a change to the fabrication process, possibly to increase yield, but after the technology has been qualified for a radiation application, which then affects the technology's radiation tolerance. Communication with the vendor can be most helpful in this regard but if long delays occur between the qualification and production, some re-qualification tests may be in order as well as ongoing quality assurance testing of production units.

One last note about radiation damage is that no electronic component, for the most part no material in general, is completely immune to radiation damage. The term "radiation hard" often gives the impression of complete immunity to radiation damage, but that is not the case. Radiation immunity always includes a qualifier as to what level of total dose and total fluence the parts are immune to damage with possible other qualifiers such as dose rate and SEU cross-section (probability).

*VI-b Low Noise*

As mentioned earlier, signals from typical particle sensors can be in the range of fCs. As an example, the most probable signal from a minimum ionizing particle passing through 300 μm of silicon is roughly 3.5 fC, with a Landau distribution (A Landau distribution is similar to a Gaussian but is asymmetric about its peak with a longer tail on the high end and closely resembles the probability of energy loss in material [21,22].) such that a threshold for signal detection must be no larger than 0.75 to 1 fC in order to assure acceptable detection efficiency. The noise usually presents a Gaussian distribution. Further taking the example of the hit/no-hit readout of Figure 9, the noise occupancy (the probability of noise mimicking a hit signal) is related directly to the relative size of the noise signal compared to the threshold setting of the comparator as given by:

$$Noise\ Occupancy \propto \exp-(\frac{V_{threshold}^2}{2\sigma^2})$$

where $V_{threshold}$ is the threshold setting of the comparator and σ is the standard deviation of the Gaussian noise distribution. It is important that this noise occupancy be much less than the expected occupancy of real hits. For the ATLAS SCT detector, the expected occupancy of real hits is ~1%. Therefore, a noise occupancy specification of $5 \times 10^{-4}$ was established. Setting the noise occupancy in the above equation to that value and solving for ($V_{threshold}/\sigma$) results in a signal-to-noise requirement ($V_{threshold}/\sigma$) of 4-to-1. It is typical for pixel detectors to set much more aggressive signal-to-noise ratios of something like 20 or 25-to-1. This is both possible because of the much smaller capacitance of the single pixel cell and desirable because the lower noise occupancy reduces confusion of noise hits from real hits in the region of higher particle densities closer to the interaction point where pixel detectors typically reside. For accurate analog measurements, for example the energy deposited in a calorimeter, the constraints on the electronic noise may be even more severe. Twelve-bit or even 16-bit ADCs are now possible for on-detector ICs, implying least-bit accuracy of 1 out of four thousand to sixty thousand. Such a dynamic range is often desirable to match the large range of energies of particles. An adequate dynamic range for the analog circuitry is then required and the noise should nominally be no more than half the least significant bit value. A complete discussion of electronic noise can be found in [24].

A simple form of analog readout is sometimes employed when only a few bits are required. This time-over-threshold (TOT) technique starts with the basic hit/no-hit circuit shown in Figure 9 and provides circuitry at the output of the comparator block that will measure the amount of time the comparator output remains above threshold. Depending upon the linearity of the circuitry, this can be a crude measurement of the amount of charge deposited. This can be simply implemented with a counter enabled by the output of the comparator. The circuitry is often enhanced by distorting the function of the shaper circuit to stretch the signal rather than shorten it to fit into a natural time interval of the sensor. Care must be taken in analyzing the effective noise of the output bit count of such a circuit as a slowly falling trailing edge of the comparator output can result in somewhat larger effective noise than expected [15].

The capacitance of the sensor device is a critical parameter for the design of low noise readout electronics. The capacitance of silicon sensors can range from a few hundred femto-Farads (fF) for small pixels (e.g. with a pixel area of tens of microns by a few hundred microns) to a few hundred pico-Farads (pF) for micro-strip sensors many centimeters long. Other more macroscopic devices like a calorimeter or wire chamber can have capacitances in the nano-Farad (nF) range.

The other critical parameter is the time available for sampling the sensor signal. Time is required to collect the charge released in the sensor. For silicon sensors this is of order 10 ns. For larger devices like the LAr electrode or a wire chamber it can be 100s of nanoseconds. In addition to consideration of charge collection time, depending upon the type of noise present, longer shaping times normally result in lower noise as the oscillating noise signal is averaged out. But the time between the possible arrivals of separate particles also constrains the time available for processing. The beam structure of the LHC, for example, collides bunches of protons every 25 ns. If the detector is going to separate particles coming from consecutive beam crossings, the electronics must be able to isolate signals within that time interval. Other machines such as the electron-positron collider for the BaBar detector had a much slower repetition rate, which allowed a much longer integration time. The electronics for the ATLAS SCT detector uses a peaking time of 20 ns.

The transconductance, which affects the input impedance, of the first stage is critical to good noise performance. Bipolar devices with low input base resistance have been successful at achieving acceptable noise levels for applications like the Liquid Argon Calorimeter and many silicon micro-strip detectors requiring short shaping times [14,16,17]. With longer shaping times (of order 100 ns or more) or with lower capacitances like pixel sensors, CMOS front-end circuits have been quite successful [10,18]. The book by H. Spieler [2] includes a detailed methodology for working with these constraints of load and timing.

*VI-c Low Power*

Power dissipation of the electronics is a serious constraint. Bringing power into the detector elements requires cabling. Removing heat generated by that power requires plumbing for some type of coolant. Both of these services require holes somewhere in the coverage of the detector for entry points. Furthermore, routing of these services through the active area of the detector creates scattering material for particles being studied. The tracking detectors such as the pixel and silicon micro-strip detectors measure the curvature of charged tracks through a magnetic field. When the particles encounter extra material such as cabling or plumbing, they have a probability to scatter, altering their path through the magnetic field and thus creating an error in the momentum measurement. Also, energy lost in traversing material before reaching one of the calorimeters will interfere with accurate measurement of the particles' initial energy. For all of these reasons, material inside the active volume of a detector must be kept to a minimum and that includes material for services.

In fact, the on-detector electronics also represents extra material, which must be accounted for, however, the alternative of sending raw sensor signals outside the detector volume would incur tremendous noise penalties so it is far preferable to locate readout circuits immediately at the sensors and include some data compaction functionality so as to minimize the amount of cabling for transmitting signals off detector. All on-detector components, however, are included in the material count and must be kept to a minimum, especially for example large capacitors or power devices. Also, for the readout circuitry of internal detectors (e.g. pixel and silicon strip detectors), bare chips are either bump bonded or wire bonded to sensors and wire bonded directly onto hybrid interconnect circuits. Commercially packaged parts take up too much space and add material. Figure 14 demonstrates how a compact, low mass silicon micro-strip detector can be built with unpackaged ICs on a polyimide/copper interconnect hybrid with a carbon fiber substrate. The term "interconnect hybrid" is used in this case for a printed and etched circuit to connect devices of different technologies. The two sensors on the front side are paired with two more on the backside rotated by a small angle to provide a stereo view of two dimensions. A piece of thermal pyrolytic graphite (TPG) is sandwiched between the two layers of sensors to extract heat. Beryllium-oxide (BeO) facings provide a heat path to a cooling pipe not shown [3].

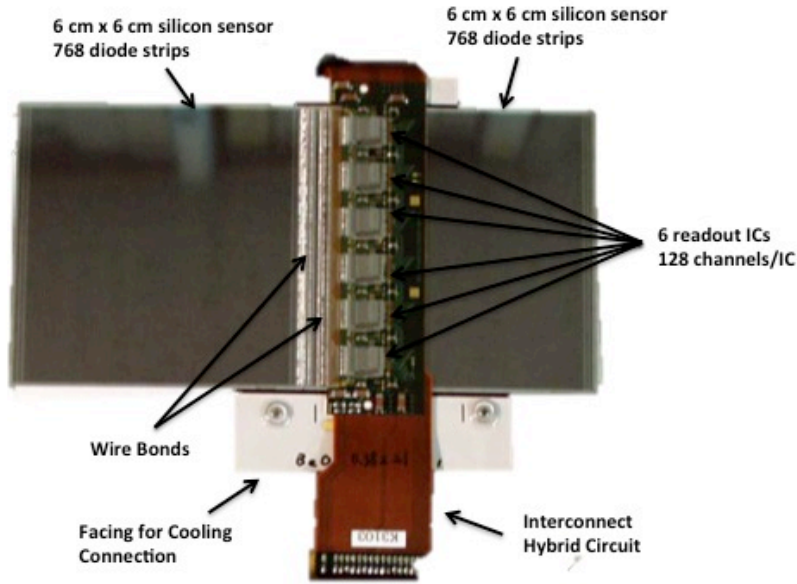

**Figure 14: An ATLAS SCT Detector Module**
**Visible are two 6 cm x 6 cm silicon micro-strip sensors and six readout ICs on an interconnect hybrid. Two more sensors are on the backside rotated by a small angle to provide a stereo measurement. The interconnect hybrid wraps around to the backside with six more readout ICs.**

This requirement for lower power adds further challenges to the low noise, fast integration time requirements already discussed. Again, the continuing efforts of the commercial electronics markets support the advances in detector technology making the present and future experiments possible. As examples of what has been achieved in this regard, Table 2 shows the power per channel and total power of the two existing ATLAS silicon detectors. Note the large channel counts, both for each detector system and for each IC. These large channel counts per IC are typical in order to minimize the material and the power consumption of the readout electronics and are another example of the advantages made from advances in circuit integration.

**Table 2: Power Consumption of the ATLAS Silicon Detector Systems (from [8,12])**
**(does not include power for optical data transmission or monitoring functions)**

|  | Power/channel | Channels/IC | Power/IC | Total # of Channels | Total Power |
|---|---|---|---|---|---|
| ATLAS Pixels | 100 µW | 2880 | 288 mW | $80 \times 10^6$ | 8 kW |
| SCT | 3 mW | 128 | 400 mW | $6.3 \times 10^6$ | 19 kW |

The ICs represented in Table 2 were designed over 10 years ago and no longer represent present state of the art technologies. The ATLAS Pixel ICs were built on a 250 nm commercial CMOS technology and the SCT ICs on an 800 nm BiCMOS technology that was designed to be radiation hard. Indeed, the next generation of this detector is now being developed on more advanced commercial technologies with a target to reduce the power consumption much further as the channel count will increase by roughly a factor of 10.

As the technologies shrink in size, the general expectation is that power consumption will also decrease, but this is not an automatic conclusion. Transistor bias voltages are reaching their limit to provide useful circuits, especially analog circuits. Some reduction in digital switching currents is anticipated but this also increases the susceptibility for SEU errors.

*VI-d High Data Rates*

As raised in section IV-d above, today's detectors, especially the ones used in colliders, produce large amounts of data that must be transmitted to the off-detector electronics and computers in short amounts of time. The ATLAS detector, for example, generates 130 gigabytes of data every millisecond. To solve this problem optical transmission has been adopted, however, radiation hardness and low power requirements have prevented some of the most cutting-edge technologies from being used. Presently operating detectors, again designed more than a decade ago, run their optical links at 40-50 Mbits/s. The next generation, however, will need to resort to higher speeds. Gigabit/s links are now in development, which will need to meet the radiation requirements including acceptable error rates. Given the planned bandwidths of multiple gigabits/s, a single event upset in the transmission line will likely cause not one bit in error but a whole burst of error bits. For this reason, error correction encoding to not only detect but fix bursts of errors will be required. Such burst error correcting techniques are well established in industry. The detector designers will need to analyze the possible error patterns when their electronics are exposed to the expected radiation and then choose the appropriate error correcting encoding scheme.

*VI-e High Reliability*

Since the advent of solid-state technologies, the ICs incorporated into particle detectors have usually been the most reliable component. The most common failures in the electronics have been connectors. They fail for many reasons: insufficient strain relief causes broken wires at the connector when the cable is pulled by accident or intentionally, multiple insertions during testing and integration wear-out the metal contacts, corrosion of the metal contacts, improper assembly of the connector. Quite often cabling and connectors are the last thing designers think about because they are not very glamorous. This results in insufficient design effort and subsequent poor reliability. As a general rule, minimizing the number of connectors will improve reliability and budgeting sufficient design effort on the cables and connectors will be rewarded with better reliability.

Other electronics, primarily solid-state devices, have earned a reputation for high reliability. We have grown accustomed to what is called the "bathtub curve" for the failure rate of solid-state devices versus time. This means that there is a relatively modest rate of failures when new devices are first turned on, typically caused by some slight imperfections in the fabrication process, followed by a long period with an extremely low failure rate, ending eventually with a steep rise in the failure rate as the devices reach their end of life. The period of extremely low failure rate has typically been many, many years. This has led to the practice of "burning in" new devices in order to find and discard those that will fail early by running them for a short time at elevated temperature and possibly voltage. These early failures are referred to as "infant mortality" failures. If you contrast this with the typical failure rate of a mechanical device, for example a motor, a practice of burning in the motor would only shorten its lifetime because a motor has definite wear-out mechanisms, which the burn-in process would exercise.

Recently, some concerns have arisen that this "bathtub curve" may no longer be the correct model for IC lifetimes. There have been some data indicating that after the period of infant mortality, the failure rate does not remain flat at an extremely low value but rather has a small positive slope indicating some wear-out mechanisms exist. The possible wear-out mechanisms so far identified include: electro-migration, time dependent dielectric breakdown, negative bias temperature instability, hot carrier injection. For more discussion of these topics see the paper by A. Hava et al. [20]. These mechanisms have been identified for a long time but they have become more of a concern as the device feature sizes have reached sub-micron dimensions. A related concern is that as commercial manufacturers continue to push performance and cost, long-term reliability may not be watched as carefully as in the past. Since many commercial products like computers, smart phones, electronic games become outdated and replaced every 2-3 years, the IC manufacturers may not be concerned about lifetimes of 10 years or more. The resulting lesson is that designers of future detectors should study the lifetimes of the IC technologies they intend to use. In many cases, these wear out mechanisms can be avoided by making the IC designs more relaxed, e.g. electro-migration is the wear-out of conductors due to high current density. By increasing conductor widths in the ICs to wider values than called for in the design rules, the problem can be reduced at the expense of possibly less dense circuitry in the IC. At the end, lifetime tests should be performed on as many components as possible, especially for components which do not have published industry lifetime test results.

Another way to improve reliability is to provide redundancy, that is spare components installed in the detector such that they can be switched on if their partner components die. If it were possible, there would be full redundancy for all components but that would be too costly in volume, material and money. Instead, all parts of the system should be analyzed for possible single point failures. Wherever possible these can be eliminated by adding redundant components. In some cases, like a data link, mechanisms can be provided whereby two units, which have their own links, can share a link if one is lost without a significant loss of performance. This was the scheme adopted by the ATLAS Semi-conductor Tracker (SCT).

The other aspect of these detectors that affords better overall lifetimes is that it is not necessary for absolutely every channel to be operational for the full detector to work satisfactorily. For example, the tracking part of the detector must find the trajectories in the magnetic field of all charged particles emanating from a scattering. In the best of all cases, that might only require hits (signals) at three separate points, however, given the possibility for scattering inside the tracking volume as well as the decay of a particle in flight, trackers are designed with many more layers. This affords a type of redundancy. If one channel on a particular layer should fail, there are normally sufficient remaining channels alive along the trajectory of each particle's path for it to be reconstructed and thereby its momentum determined. In some cases, such a dead channel may increase the ambiguity for a track, which suddenly veers off, but this can be handled by statistics of many, many events. Calorimeters operate in a similar fashion by sampling both the lateral and longitudinal extent of the energy deposited. Of the millions of channels in each sub-detector, it is normally acceptable for a few percent of channels to be dead and this is verified with simulation studies before the detector is ever built. This allowed number of dead channels not only extends the useful life of the detector but actually makes the detector buildable. An absolutely perfect detector (i.e. no dead channels) with hundreds of millions of channels is probably not possible to build.

## VII Doing Physics with a Particle Detector

It is far beyond scope of this Primer to review all the different ways to construct a high energy particle physics experiment using the types of particle detectors and readout electronics discussed here. Even the design of one such experiment would include a discussion of the physics to be investigated, methods to create the desired reaction to study, possible backgrounds, the necessary size of the event sample, the data acquisition system, the proposed analysis procedure and required analysis resources. However, given the expected audience for this Primer, it might be useful to at least give a taste of how these detectors are typically employed.

One of the primary goals of such an experiment is the search for new particles. Beyond that, experiments attempt to measure characteristics of such new particles, for example, the probability to produce them, typically called the cross section, the lifetime of the particle, typically measured as the uncertainty of the particle's mass for very short-lived particles, and a list of its decay products and relative probabilities to decay into each product. This information does not just add to the inventory of discovered sub-atomic particles. It also gives information about the dynamics of how these particles interact with each other and thereby about how matter in the universe is structured, possibly even how the universe has evolved.

Stepping back from such profound concepts as how the universe evolved, we can look at an example of how a new particle can be identified. In sections II & III, we saw how the tracks charge particles leave in detectors can be used to measure the particles' momenta and their charge signs when those detectors are placed in a strong magnetic field, and how neutral particles can be identified by their lack of tracks in tracking detectors but how their energy can be measured in calorimeters. The basis of the analysis of any experiment then is to find all the tracks of each interaction, measure the charge and momentum of each, and measure the energy of all neutral particles by what each deposits in calorimeters. A few techniques can also be applied to identify the type of as many of the particles as possible. For example, electrons and positrons are charged particles that behave very much like neutral gammas when they enter a calorimeter while their heavier cousin, the muon, deposits very little energy in a calorimeter and can be detected exiting the back of the calorimeters. Charged pions of similar mass to muons are typically stopped by calorimeters but deposit their energy in a more localized region by nuclear interactions rather than spread along a long path like electrons, positrons and gammas due to their sole electromagnetic interactions. Then each particle type has a typical signature.

Given all this information, there are a few ways to discover a new particle. One of the most straight-forward is to search for the possible mass of a particle that decayed into the observed known particles. An example of this was used in the early running of the ATLAS experiment when beam intensity was very low, not to discover a new particle yet, but rather to use the technique to check the accuracy of the tracking part of the ATLAS detector. To start, in a large number of proton-proton inelastic scattering events, all of the identified charged pion particles were combined as pairs of a $\pi^+$ and a $\pi^-$ particle. For each pair, the hypothetical mass of a particle that decayed into that pair of pions was calculated and put on a plot, a mass plot. (The calculation applies conservation of relativistic energy and momentum to the reaction of a particle of unknown mass decaying into two particles of known mass and measured momentum, then solving for the unknown mass. It's just a bit of special relativity calculations.) That plot is shown in Figure 15. A very distinct peak appears in the plot, a Gaussian distribution. The pion pairs whose momenta form a hypothetical mass in that peak likely decayed from a particle with a mass at the center of that peak. The pion pairs whose momenta form a hypothetical mass in the flat background either came from some other part of the interaction or they were just paired with the wrong other pion in their specific calculation. The particle with a mass matched to that peak is called a K-short meson ($K^0_s$) that was identified in 1947.

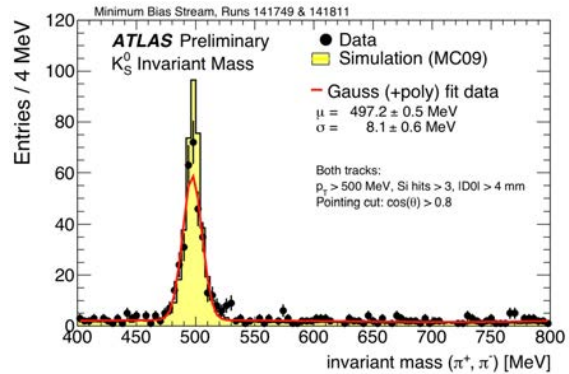

Figure 15: Mass Plot with $K_s$ Peak [25]

The reason this was interesting for the ATLAS experiment is that the accuracy and precision of the measured peak and width of that mass distribution were a very good check of the accuracy of the ATLAS tracking detector to measure track momenta, e.g. the magnetic field was very accurately understood. Figure 16 is a computer reconstruction of one event in which the reconstructed mass of the two pions sits in the $K^0_s$ mass peak and the tracks of those two pions are highlighted in red. Even though the track data is made up of discrete points in space from silicon detectors and not a continuous string of droplets in a cloud chamber or bubbles in a bubble chamber, there are a sufficient number of silicon detectors recording 3-coordinate positions for a computer to reconstruct very precise track positions just like the picture in Figure 2.

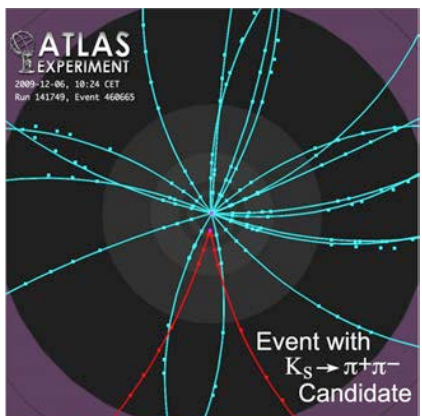

Figure 16: ATLAS Display of a $K_s$ Reconstructed Event [25]

A very similar technique was used by ATLAS in 2012 to identify and discover the long sought Higgs particle. The theory of the Higgs field predicted this particle and also predicted what other particles it was likely to decay into but did not predict its mass. The search for the Higgs involved analyzing all the data being produced by the ATLAS and the CMS detectors, calculating hypothetical masses of all combinations of particles predicted by the theory to be decay products of the Higgs. This was an enormous job requiring many, many physicists and many high-speed computers. A big complication to the search was that the probability to produce a Higgs was believed to be very low, therefore, the experiment needed to create a large number of interactions. To accomplish that in a finite amount of time (a small number of years of data collection) the colliding beam intensity had to be high enough to produce collisions very frequently, so frequently that multiple interactions occur simultaneously in the detector. A typical "event", that is the crossing of two bunches of protons from the two intersecting storage rings, is shown in the right side of Figure 17. You can see at least fifteen proton-proton collisions identified by colored circles, typically there are up to 25 such collisions in each crossing of the two beams. In this particular event, the computer has identified two tracks highlighted in yellow to have come from a short-lived particle called the Z. The number of extra collisions occurring in each event as well as the number of particles produced by each collision create a formidable background to remove from any serious analysis.

One expected decay product of the Higgs was a pair of gammas and so one of the many searches for bumps in mass plots, euphemistically called "bump hunting", was the two-gamma mass plot. The top plot on the left side of Figure 17 shows such a mass plot for two gammas with a small bump sitting on top of a very fast falling background. Finding that peak on top of the background was not easy, but the lower plot shows the enhanced peak with the background understood and subtracted. That is the data analysis that prompted the announcement on July 4$^{th}$ of 2012 that a new particle had been found that was "consistent with this Higgs Boson" [26]. The announcement was a very conservative one, not claiming yet with certainty that the Higgs had been found but only that the evidence was consistent with what had been sought. It was only after analysis of much more data showed that all the characteristics of this particle at the identified mass matched the theory that a firm claim of discovery was made.

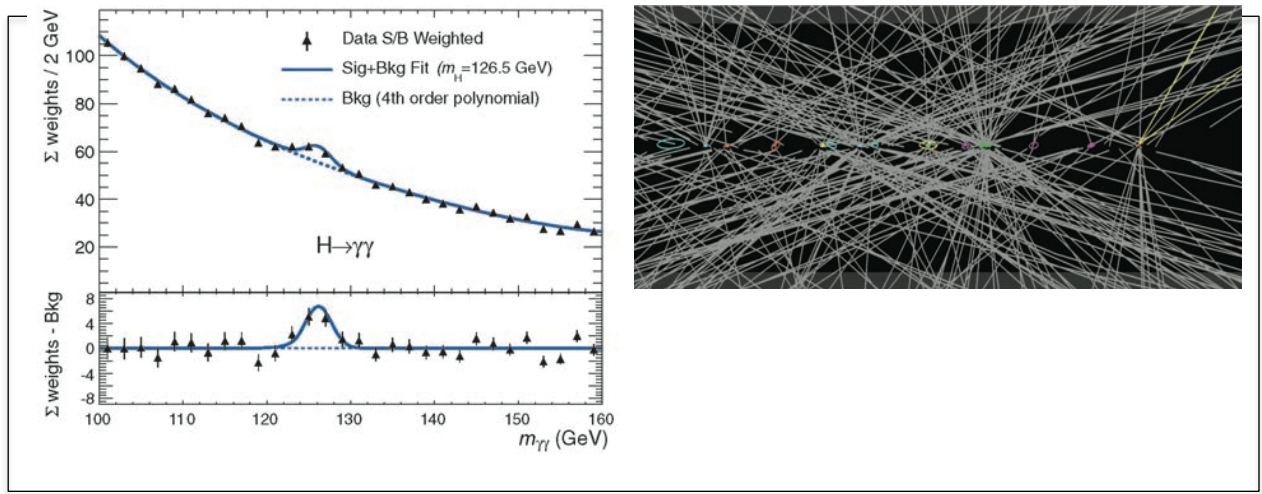

**Figure 17: Discovering New Particles in the ATLAS Detector [26]**
**Left is the mass plot of two gammas leading to the discovery of the Higgs particle.**
**Right are the reconstructed tracks of a typical ATLAS event.**
**A Z particle decaying to two muons is highlighted in yellow at the far right.**

## Acknowledgments

The author wishes to thank several SCIPP scientists who encouraged that such a paper would be a useful introduction for our students and supported it by proof reading the original version followed by encouraging their students to read it, namely Abe Seiden, Hartmut Sadrozinski, Jason Nielsen and Vitaliy Fadeyev. Helmuth Spieler's book [2] was very helpful as was Helmuth himself in allowing the use of its material and figures especially in writing the precursor of this paper, the chapter "Electronics Requirements for Collider Physics Experiments", in the book *Extreme Environment Electronic*s, J.D. Cressler & H.A. Mantooth, Ed., Boca Raton, FL, CRC Press pp 887-894 (2012). Lastly, a special thanks to our student Jennifer Volker who graciously agreed to read the original version providing many useful comments from an undergraduate student perspective allowing the revision to include more background and connections that hopefully will make the paper more understandable to its target audience.

## Appendix: Definition of Radiation Exposure Units

There are a few different measurements of radiation exposure depending upon the type of radiation and what is being exposed and potentially damaged. The exposure to ionizing radiation called dose is measured by the amount of energy that would be absorbed by a given amount of mass. The commonly used unit is a Gray defined to be 1 Joule absorbed per kilogram, abbreviated as "Gy". An older unit of dose that is no long considered standard but still used by some is a "rad" with 1 rad = 0.01 Gy.

Just for fun we can consider how much energy is absorbed by a dose of ionizing radiation and compare that to something possibly more familiar. The exposure to ionizing radiation for the ATLAS Inner Pixel Detector listed in Table 1 is 500 kGy. The mass of the Pixel detector was estimated in the technical design report to be 38.1 kg [27], which yields a total energy absorbed over the target 10-year life of the detector to be:

$$500 \text{ kGy} = 500 \text{ kJoule/kg}$$

$$\text{Energy Absorbed} = 500 \text{ kJ/kg} \times 38.1 \text{ kg (over life of the detector)}$$

$$= 19{,}050 \text{ kJ}$$

Over the 10-year life of the detector, the likely running duty cycle is estimated to be approximately 66%, therefore, the number of exposure seconds in the life of the detector will be:

$$\text{Detector lifetime} = 60 \text{ sec/min} \times 60 \text{ min/hr} \times 24 \text{ hr/day} \times 365 \text{ days/yr} \times 10 \text{ yrs} \times 66\%$$

$$= 2.08 \times 10^8 \text{ sec}$$

Therefore, the power absorbed by the Pixel detector while the LHC is operating should be about:

$$\text{Power Absorbed} = 19{,}050 \text{ kJ} / (1.89 \times 10^8 \text{ sec})$$

$$= 0.092 \text{ Watts (1 Watt} = 1 \text{ Joule/sec)}$$

The heat generated by the electronics of the detector and the sensor leakage current will be in the range of kilowatts (8kW in Table 2), much larger than this heat due to radiation absorption, and require the detector to be cooled. The cooling system, then, will more than adequately remove any heat generated by the radiation exposure. The reason the energy absorbed is so small is partly due to the mass of the silicon pixel detector being low. The actual damage done to instruments containing semiconductors can be quite severe, not because of energy absorbed, which might be turned into a small amount of heat, but because the electrical characteristics of the devices are altered, which are discussed further in sections IV and VI. The amount of energy absorbed per unit mass is just a convenient way to quantify the amount of radiation exposure. Still for most non-biological material, this measure of exposure is a useful one to scale the amount of damage to be expected.

Energy absorption for non-ionizing radiation such as neutrons is a more complex problem because these particles can travel through material without losing any energy until they hit an atomic nucleus either scattering elastically, inelastically or, if they are traveling slowly enough, being captured by the nucleus. It is basically a statistical nuclear process. The common measure of non-ionizing radiation is just the number particles passing through a unit of area called fluence and normalized to the equivalent amount of damage done by neutrons of 1 MeV energy, abbreviated as $n_{eq}/cm^2$. Exposure to particles that are both charged and subject to the nuclear strong force such as protons then result in energy deposited by both ionization and by nuclear collisions. Exposure to such particles must be quantified both in terms of the dose and the fluence of the exposure.

For biological material, especially human tissue, different units of measure are used. The Rem (Roentgen-equivalent-man) or the Sievert (Sv) takes into account the relative biological sensitivities to different forms of ionizing radiation, but this whole topic is beyond the scope of this paper. A health physics department would be a good source of information in this area.

# REFERENCES

[1]     C. Grupen, *Particle Detectors*, Cambridge, UK: Cambridge University Press, 1996.

[2]     H. Spieler, *Semiconductor Detector Systems*, Oxford, UK: Oxford University Press, 2008.

[3]     G. Aad et al., "The ATLAS Experiment at the CERN Large Hadron Collider", 2008 *JINST* 3 S08003.

[4]     P.H. Adrian et al., "Search for a dark photon in electroproduced *e*+*e*−e+e− pairs with the Heavy Photon Search experiment at JLab", Phys. Rev. D98 no.9, pp. 0911011-6, 2018.

[5]     M.Battaglieri et al., "The Heavy Photon Search Test Detector", Nucl. Instrum. Meth. A777, pp. 91-101, 2015.

[6]     S. Chatrchyan et al., "The CMS experiment at the CERN LHC", 2008 *JINST* 3 S08004.

[7]     K. Aamodt et al., "The ALICE experiment at the CERN LHC", 2008 *JINST* 3 S08002.

[8]     A. Augusto et al., "The LHCb Detector at the LHC", 2008 *JINST* 3 S08005.

[9]     B. Aubert et al., "The BABAR detector", *Nuclear Instruments and Methods in Physics Research A*, vol. 479, pp. 1-116, 2002.

[10]    G. Aad et al., "ATLAS pixel detector electronics and sensors", 2008 *JINST* 3 P07007.

[11]    Y. Arai et al., "ATLAS Muon Drift Tube Electronics", 2008 *JINST* 3 P09001.

[12]    W. Snoeys, G. Anelli, M. Campbell, E. Cantatore, F. Faccio, E.H.M. Heijne, P. Jarron, K.C. Kloukinas, A. Marchioro, P. Moreira, T. Toifl, K. Wyllie, "Integrated circuits for particle physics experiments", *IEEE Journal of Solid State Circuits,* vol. 35, no. 12, pp. 2018-2030, 2000.

[13]    T. Calin, M. Nicolaidis, R. Veazco, "Upset hardened memory design for submicron CMOS technology", *IEEE Tranactions on Nuclear Science*, vol. 43, pp. 2874-288, 1996.

[14]    F. Campabadal et al., "Design and performance of the ABCD3TA ASIC for readout of silicon strip detectors in the ATLAS semiconductor tracker", *Nuclear Instruments and Methods in Physics Research A*, vol. 552, pp. 292-328, 2005.

[15]    R. Becker, A. Grillo, R. Jacobsen, R. Johnson, I. Kipnis, M. Levi, L. Luo, P.F. Manfredi, M. Nyman, V. Re, N. Roe, S. Shapiro, "Signal processing in the front-end electronics of BaBar vertex detector", *Nuclear Instruments and Methods in Physics Research A*, vol. 377, pp. 459-464, 1996.

[16]    N.J. Buchanan et al., "ATLAS liquid argon calorimeter front end electronics", 2008 *JINST* 3 P09003.

[17]    S. Rescia, "A SiGe Front-End Prototype for the Upgraded ATLAS Liquid Argon Calorimeter", 2*009 IEEE Nuclear Science Symposium Conference Record*, pp, 1134-1137, 2009.

[18]    V. Re et al., "Performance of the BABAR silicon vertex tracker", *Nuclear Instruments and Methods in Physics Research A*, vol. 501, pp. 14-21, 2003.

[19]    M. Ullán et al., "Ionization Damage on ATLAS-SCT Front-End Electronics Considering Low-Dose-Rate Effects", *IEEE Transaction on Nuclear Science*, 49, 1106, 2002.

[20]     A. Hava et al., "Integrated Circuit Reliability Prediction Based on Physics-of-Failure Models in Conjunction With Field Study", *IEEE Reliability and Maintenance Symposium (RAMS), 2013 Proceedings*, 2013.

[21]     L. Landau, "On the Energy Loss of Fast Particles by Ionization", J.Phys. USSR 8, p. 201, 1944.

[22]     D.H. Wilkinson, "Ionization energy loss by charged particles Part I. The Landau distribution", *Nuclear Instruments and Methods in Physics Research A*, vol. 383, pp. 513-515, 1996.

[23]     A. Ahmad et al., "The silicon microstrip sensors of the ATLAS semiconductor tracker", Nuclear Instruments and Methods in Physics Research A, vol. 578, pp. 98-118, 2007.

[24]     P. Horowitz and W. Hill, *The Art of Electronics*, Cambridge, UK: Cambridge University Press, 1989.

[25]     G. Aad, Kshort and Λ production in *pp* interactions at $s\surd$=0.9 and 7 TeV measured with the ATLAS detector at the LHC, Phys. Rev. D85. pp. 012001-16, 2012.

[26]     The ATLAS Collaboration, "A Particle Consistent with the Higgs Boson Observed with the ATLAS Detector at the Large Hadron Collider", Science 338, pp. 1576-1582 2012.

[27]     N. Wermes, *ATLAS pixel detector: Technical Design Report*, CERN-LHCC-98-013; ATLAS-TDR-11, ISBN 9290831278, 1998.